\documentclass[a4paper,fleqn,usenatbib]{mnras}
\usepackage{amssymb,amsmath,graphicx,psfrag,mathtools}
\usepackage[T1]{fontenc} 
\usepackage{ae,aecompl} 
\usepackage{url}
\usepackage{revsymb}
\hypersetup{draft} 
\title[The FRB-SGR Connection]{The FRB-SGR Connection}
\author[J. I. Katz]{
J. I. Katz,$^{1}$\thanks{E-mail katz@wuphys.wustl.edu} 
\\
$^{1}$Department of Physics and McDonnell Center for the Space Sciences,
Washington University, St. Louis, Mo. 63130 USA 
}
\date{Accepted XXX.  Received YYY; in original form ZZZ} 
\pubyear{2020} 
\date{\today}
\begin{document} 
\label{firstpage} 
\pagerange{\pageref{firstpage}--\pageref{lastpage}} 
\maketitle 
\begin{abstract}
	The discovery that the Galactic SGR 1935$+$2154 emitted FRB 200428
	simultaneous with a gamma-ray flare demonstrated the common source
	and association of these phenomena.  If FRB radio emission is the
	result of coherent curvature radiation, the net charge of the
	radiating ``bunches'' or waves may be inferred from the radiated
	fields, independent of the mechanism by which the bunches are
	produced.  A statistical argument indicates that the radiating
	bunches must have a Lorentz factor $\gtrapprox 10$.  The observed
	radiation frequencies indicate that their phase velocity (pattern
	speed) corresponds to Lorentz factors $\gtrapprox 100$.  Coulomb
	repulsion implies that the electrons making up these bunches have
	yet larger Lorentz factors, limited by their incoherent curvature
	radiation.  These electrons also Compton scatter the soft gamma-rays
	of the SGR.  In FRB 200428 the power they radiated coherently at
	radio frequencies exceeded that of Compton scattering, but in more
	luminous SGR outbursts Compton scattering dominates, precluding the
	acceleration of energetic electrons.  This explains the absence of a
	FRB associated with the giant 27 December 2004 outburst of SGR
	1806$-$20.  SGR with luminosity $\gtrsim 10^{42}$ ergs/s are
	predicted not to emit FRB, while those of lesser luminosity can do
	so.  ``Superbursts'' like FRB 200428 are produced when narrowly
	collimated FRB are aligned with the line of sight; they are unusual,
	but not rare, and ``cosmological'' FRB may be superbursts.
\end{abstract}
\section{Introduction}
Soft Gamma Repeaters (SGR) have long been candidates for the sources of Fast
Radio Bursts (FRB).  SGR are believed to originate in young neutron stars
with extremely high magnetic fields and to be powered by dissipation of
their magnetostatic energy \citep{K82,TD92,TD95}, offering an ample source
of energy.  The energies $\sim 10^{40}$ ergs of even ``cosmological'' FRB are
a tiny fraction of the $\sim 10^{47}$ ergs of magnetostatic energy of a
neutron star with a $\sim 10^{15}$ gauss field, a value inferred from the
spindown rates of some SGR, measured in their quiescent Anomalous X-ray
Pulsar (AXP) phases.

SGR also have short characteristic time scales.  The most intense parts of
their outbursts typically last $\sim 0.1$ s, {as did the outburst of SGR
1935$+$2154 simultaneous with FRB 200428.  Upper bounds on the rise times of
the three giant Galactic SGR outbursts were $< 1$ ms \citep{K16}}.
Although the temporal structure of SGR have not been measured on the scale
of the fastest temporal structure of FRB ($\sim 10\,\mu$s \citep{Ch20}), the
fact that both display extremely short time scales, shorter
than any other astronomical time scale except those of pulsar pulses,
suggests an association.  This hypothesis has been advanced by many authors
\citep{C16,CW16,D16,K16,Z17,W18,WT19}; see \citet{K18a} for a review.

The discovery of FRB 200428 in association with an outburst of SGR
1935$+$2154 demonstrated that FRB and SGR can be produced by the same events
but raises four questions:
\begin{enumerate}
	\item Why is SGR 1935$+$2154 different from other SGR, producing an
		observable FRB when the much more powerful 27 December 2004
		giant eruption of SGR 1806$-$20 did not?
	\item What are the parameters of the emitters?
	\item Why do SGR 1935$+$2154 bursts range by $\ge 8$ orders of
		magnitude in $F_{radio}/F_\gamma$?
	\item What predictions can we make?
\end{enumerate}
\section{The Problem}
FRB 200428, with fluence $\sim 1.5 \times 10^6\,$Jy-ms was discovered by
CHIME/FRB \citep{C20} and by STARE2 \citep{B20} during an outburst of the
SGR 1935$+$2154\footnote{Sometimes referred to as J1935$+$2154.} observed by
INTEGRAL \citep{M20}, Insight-HXMT \citep{Li20}, Konus-Wind \citep{R20} and
AGILE \citep{T20} and consistent with the location of the SGR.  A burst
detected two days later by FAST \citep{FAST} had a fluence of
$\sim 60\,$mJy-ms, a factor $\sim 4 \times 10^{-8}$ times that of FRB
200428, while \citet{Ki20} found two bursts of fluences $\sim 112\,$Jy-ms
and $\sim 24\,$Jy-ms, about 1.4 s apart, 26 days after FRB 200428.

The ratio of the STARE2 \citep{B20} radio to the Insight-HXMT soft gamma-ray
\citep{Li20} fluences of SGR 200428 was $\sim 2 \times 10^{12}$
Jy-ms/(erg/cm$^2$).  This was several orders of magnitude greater than
the upper limit of $10^7$ Jy-ms/(erg/cm$^2$) set by \citet{TKP16} on any FRB
associated with the giant 27 December 2004 outburst of SGR 1806$-$20.

The large observed radio-frequency fluence \citep{B20} of FRB 200428, taking
a distance of 6 kpc, a compromise among the 12.5 kpc \citep{KSGR18}, 9.1 kpc
\citep{Zh20} and 6.6 kpc \citep{ZZ20} estimated for the embedding SNR
G57.2$+$0.8 and the 2--7 kpc estimated by \citet{M20} from dust-scattered
SGR emission, implies an isotropic-equivalent emitted energy $\sim 10^{-6}$
that of a nominal 1 Jy-ms ``cosmological'' FRB at $z = 1$.  Any explanation
of FRB as products of SGR must be consistent with ``cosmological'' FRB whose
radio emission is several orders of magnitude more energetic than that of
FRB 200428 and with the radio-to-gamma ray fluence ratio of FRB 200428 more
than five orders of magnitude greater than that of SGR 1806$-$20.  A number
of theoretical interpretations have been suggested
\citep{LKZ20,LP20,MBSM20,WXC20,W20}.

A past argument \citep{K20} against a neutron star origin of FRB was the
absence of periodicity in repeating FRB, particularly in the well-studied
FRB 121102 \citep{Z18}.  SGR 1935$+$2154 has a period of 3.245 s
\citep{I16}, which would be expected to modulate the observable activity of
FRB 200428, whatever its mechanism of emission, unless its magnetic field be
a dipole aligned with the spin axis.  The few extant detections of FRB
200428 are insufficient to test this prediction.

{ An additional argument
\citep{K20} was that the FRB sky was not dominated by Galactic FRB, and that
therefore their sources must not be distributed with the stellar population
whose inverse distance squared-weighted mass distribution and visible and
other radiation are dominated by the Galactic disc.  The 1.5 MJy-ms fluence
of FRB 200428, about two orders of magnitude greater than the cumulative
fluence of all other observed FRB, now invalidates that argument.}
\section{The Host}
The characteristic spindown age of SGR 1935$+$2154 was measured over about
120 days in 2014 to be 3600 y \citep{I16}, several times shorter than the
estimated age of SNR G57.2$+$0.8 \citep{KSGR18,ZZ20} in which it is
embedded.  These values of the SNR age were inferred from larger estimates
of its distance; the smaller distance values of \citet{M20} and \citet{ZZ20}
would lead to much lower values of the SNR's age and might resolve the
disagreement.

Alternatively, the neutron star might now be in a period (at least several
years long because the spindown was measured six years before the FRB) of
unusual activity and unusually rapid spindown.   Yet other alternatives
include misidentification of the SGR with the SNR or the emergence of strong
magnetic fields long after the neutron star's birth \citep{B19}. 
\section{Curvature Radiation}
\label{radcharge}
FRB emission by a strongly magnetized neutron star has been explained as
coherent curvature radiation \citep{KLB17}.  Its spectrum is the product of
the spectrum of radiation emitted by accelerated point charges and the
spectrum of the spatial structure of the coherent charge density
distribution \citep{K18b}.  The spectrum emitted by an accelerated point
charge is very smooth and broad, so the observed spectral structure must be
attributed to the distribution of charge density.  The frequency and
spectrum of the emitted radiation is determined by the phase velocity
(pattern speed) of the deviations from charge neutrality that radiate.  This
must be distinguished from the velocities of the individual charges that
also radiate incoherently.  Describing the phase velocity of the plasma wave
that bunches the charge density by its corresponding Lorentz factor
$\gamma_w$, its minimum value $\gamma_{min}$ for observed curvature
radiation of angular frequency $\omega$
\begin{equation}
	\label{gamma}
	\gamma_{min} \approx \left({3 \omega R \over c}\right)^{1/3},
\end{equation}
where $R$ is the radius of curvature of the guiding magnetic field line.

We have no direct evidence that the observed radiation is near this peak of
the spectral envelope of curvature radiation, but selection effects favor
the detection of the brightest radiation and make that plausible.  This is
the same argument that justifies the assumption of particle-field
equipartition in incoherent synchrotron sources: the most efficient
radiators are the most detectable.  Taking $R \sim 10^6$ cm, the neutron
star radius, because the available energy density decreases rapidly with
increasing distance from the neutron star, leads to an estimate
$\gamma_{min} \approx 100$, only weakly dependent on the uncertain
parameters.

The observed, comparatively narrow but varying, spectral bands of FRB
radiation imply that there are comparatively few charge ``bunches''
radiating at any one time.  If there were $\gg \omega/\Delta\omega \sim
10$ such bunches, where $\Delta\omega$ is the width of an individual band,
each would likely have a different peak frequency of radiation corresponding
to a peak in the Fourier transform of the spatial distribution of charge.
The total spectrum of radiation, a sum over many such peaks, would be smooth
and broad, rather than being confined to a few narrower bands as observed.
\subsection{Radiating Charges}
\label{radiating}
We model this distribution of charge density as a single charge $Q$, the
amplitude of the peak of that Fourier transform; a actual point charge $Q$
would radiate a very broad and smooth spectrum, not seen.  The 
frequency-integrated power received per unit solid angle $\cal F$
\citep{RL79}
\begin{equation}
	\label{dPdO}
	{\cal F} = {dP \over d\Omega} = {4 Q^2 a_\perp^2 \over \pi c^3}
	\gamma_w^8 {1 - 2\gamma_w^2\theta^2\cos{2\phi} + \gamma_w^4\theta^4
	\over (1 + \gamma_w^2\theta^2)^6},
\end{equation}
where $a_\perp \approx c^2/R$ is the magnitude of the acceleration
perpendicular to the velocity (and magnetic field line), $\theta$ is the
angle between the direction of observation and the velocity vector and
$\phi$ is an azimuthal angle.  The half-width at half power of the radiation
pattern $\theta_{1/2} \approx 0.35/\gamma_w$.  For $\gamma_w\theta
\gg 1$ the final factor varies $\propto (\gamma_w\theta)^{-8}$,
cancelling the factor of $\gamma_w^8$, leading to a result independent
of $\gamma_w$ but $\propto \theta^{-8}$.  Taking $\gamma_w\theta \ll 1$ and
Eq.~\ref{gamma}, if $\gamma_w = \gamma_{min}$
\begin{equation}
	Q \approx 0.2 {c^{5/6} \over R^{1/3} \omega^{4/3}} \sqrt{dP \over
	d\Omega} \approx 5 \times 10^{-8} \sqrt{dP \over d\Omega}
\end{equation}
in Gaussian cgs units for L-band radiation.

If $\gamma_w \gg \gamma_{min}$ then the spectral peak and most of the
radiated power is at frequencies above the observed L-band.  As a result of
integrating
\begin{equation}
	\int\!{dP \over d\Omega\,d\omega}d\omega \propto
	\int_0^{\omega_{max}}\!\omega^{1/3}\,d\omega \propto
	\omega_{max}^{4/3} \propto \gamma_w^4
\end{equation}
up to $\omega_{max} \sim c \gamma_w^3/(3R)$, the inferred spectrally
integrated $dP/d\Omega$ is multiplied by $(\gamma_w/\gamma_{min})^4$.
{Here we have taken the familiar angle-averaged result (\citet{RL79}
Eq.~6.32) because the radiating charge distribution is likely to be very
oblate (Sec.~\ref{epart}) and may contain multiple incoherently adding
``bunches'', so that an observer receives a radiation from a distribution of
emitters whose velocities are spread over an angular width $\sim 1/\gamma$.
Then} Eq.~\ref{dPdO} is replaced by
\begin{equation}
	\label{dPdO2}
	{\cal F}_{obs} = \left.{dP \over d\Omega}\right|_{obs} = {4 Q^2
	a_\perp^2 \over \pi c^3} \gamma_w^4 \gamma_{min}^4 {1 - 2\gamma_w^2
	\theta^2 \cos{2\phi} + \gamma_w^4\theta^4 \over
	(1 + \gamma_w^2\theta^2)^6},
\end{equation}
where $(dP/d\Omega)|_{obs}$ is the measured power density at the 
observational frequency, henceforth 1400 MHz, corresponding
(Eq.~\ref{gamma}) to $\gamma_{min}$.

For FRB 200428 \citep{B20,C20}, taking a bandwidth of 400 MHz, a distance of
6 kpc and emission lasting 3 ms, and for a nominal ``cosmological'' FRB with
a flux density of 1 Jy at $z=1$
\begin{equation}
	\left.{dP \over d\Omega}\right|_{obs} \sim 
	\begin{cases}
		1 \times 10^{36}\,\text{erg/sterad-s} & \text{FRB 200428} \\
		2 \times 10^{42}\,\text{erg/sterad-s} & z=1
	\end{cases}
\end{equation}
and 
\begin{equation}
	\label{Q}
	Q \sim
	\begin{cases}
		5 \times 10^{10}\gamma_r^{-2}\,\text{esu} = 15\gamma_r^{-2}\,
		\text{C} & \text{FRB 200428} \\
		8 \times 10^{13}\gamma_r^{-2}\,\text{esu} = 3 \times 10^4
		\gamma_r^{-2}\,\text{C} & z=1,
	\end{cases}
\end{equation}
where $\gamma_r \equiv \gamma_w/\gamma_{min} \approx \gamma_w/100
\ge 1.$  These are only the charges whose (collimated) radiation is directly
observed.  There may be additional charges (much larger in total absolute
magnitude) radiating in other directions, either simultaneously with the
observed FRB, or at other times, if the FRB is part of a wandering or
intermittent beam \citep{K17a}.
\subsection{Empirical Lower Limit on the Lorentz Factor}
\label{Lorentz}
The upper limits set by \citet{L20} on FRB emission during other soft
gamma-ray flares of SGR 1935$+$2154 of $\lesssim 10^{-8}$ of FRB 200428
statistically constrain the Lorentz factor $\gamma_w$ of the emitting
charges (or their wave or pattern speed) if the emission is produced by
acceleration perpendicular to the velocity.  This bound applies to
synchrotron radiation as well as to curvature radiation.

For a relativistic particle of Lorentz factor $\gamma$, emission at angles
$\theta \gg 1/\gamma$ is ${\cal O}(\gamma\theta)^{-8}$ times that for
$\theta \ll 1/\gamma$ (Eq.~\ref{dPdO}).  Brightness selection effects make
it likely that FRB 200428 was observed at an angle $\theta \lesssim
\theta_{1/2} \approx 0.35/\gamma$.  If other observed soft gamma-ray bursts
of SGR 1935$+$2154 produced radio bursts intrinsically similar to FRB 200428
but beamed in directions statistically uniformly but randomly distributed on
the entire sky, then of $N$ such bursts the closest to the observer was
likely at an angle $\theta \sim \sqrt{4/N}$.  Then
\begin{equation}
	\label{gammastat}
	\gamma_w \gtrapprox 0.35 \left({{\cal F}_{max} \over
	{\cal F}_{min}}\right)^{1/8} \sqrt{N \over 4} \approx 10,
\end{equation}
where $N=29$ is the number of SGR outbursts observed by \citet{L20} and
${\cal F}_{max}/{\cal F}_{min} \sim 10^8$ is the ratio of the brightest FRB
observed (FRB 200428) to the upper limits set { on FRB emission by} all
the other SGR outbursts.  The effective (half-width at half-power) beam
width $\theta_{1/2} \approx 0.35/\gamma_w \lessapprox 2^\circ$.  Continuing
observation, increasing $N$, will either increase the lower bound of
Eq.~\ref{gammastat} or find a distribution of observed FRB strengths from
which their angular radiation pattern may be inferred.

{ Eq.~\ref{gammastat} describes the radiation of a point charge or a
collimated beam.  If the radiators are not perfectly collimated the lower
bound would be greater because the observed ${\cal F}_{max}/{\cal F}_{min}$
would be less than that describing the radiation pattern of an individual
point charge.}

{ This result is only approximate because observations are made in a
fixed frequency band, while Eq.~\ref{dPdO} represents an integral over all
frequencies.  This could be allowed for were the charge density spectrum
known (as it is for a point charge), in which case the only complication
would be the known angle-dependence of the Doppler shift.  A wave spectrum
of charge density radiates at different frequencies in different directions,
but that spectrum is not known.}

This method cannot be applied to the numerous observed bursts of FRB 121102
because no corresponding gamma-ray activity is detected { and the number
$N$ of undetected radio bursts is not known}.  Because of limits
on the sensitivity of X- and gamma-ray detectors, it is likely to be
feasible only for Galactic FRB.  { However, it is possible to relate the
observed duty factor $D$ of a repeating FRB to its intrinsic activity duty
factor $D_{activity}$ if the beam of half-width $\approx 0.35/\gamma_w$ is
isotropically and randomly distributed on the sky and it is assumed that the
observer must be within the beam half-power width to observe a burst.  Then
\begin{equation}
	D \approx {\pi (0.35/\gamma_w)^2 \over 4 \pi} D_{activity} \approx
	0.03 {D_{activity} \over \gamma_w^2}.
\end{equation}
Because $D_{activity} \le 1$, this relation can be inverted to obtain the
bound
\begin{equation}
	\gamma_w \lessapprox \sqrt{0.03 \over D}.
\end{equation}
For known repeating FRB this bound is typically $\gamma_w \lessapprox 100$.}

{Subsequent to the submission of the original version of this paper,
\citet{Y20} reported a number $N \gg 29$ of outbursts of SGR 1935$+$2154
that were not accompanied by FRB outbursts, but with upper limits
${\cal F}_{min}$ (Eq.~\ref{gammastat}) much greater than those of
\citet{L20}.  This is consistent with the extreme collimation of FRB
emission implied by Eq.~\ref{dPdO}.}
\subsection{Particle Energies}
\label{epart}
The requirement that the electrostatic repulsion of the charge bunches not
disrupt them sets a lower bound on the particle energy $E_e$ and Lorentz
factor $\gamma_{part}$; an electron must have sufficient kinetic energy to
overcome repulsion by the net bunch charge $Q$.  Coherent emission requires
that the charge bunch extend over a length $\lesssim \lambdabar = c/\omega =
\lambda/2\pi$ in its direction of motion and radiation in order that fields
from its leading and trailing edges, arriving at times separated by
$\lesssim \lambdabar/c$, add coherently.  The minimum electron energy is
\begin{equation}
	\label{Ee}
	E_e = \gamma_{part} m_e c^2 \gtrsim {Qe \over \ell},
\end{equation}
where $\ell$ is approximately the largest dimension of the charge cloud.  If
the cloud is roughly spherical $\ell \sim \lambdabar$ (about 3 cm for L-band
radiation)
\begin{equation}
	\label{lambdabar}
	E_e \ge 
	\begin{cases}
		5\gamma_r^{-2}\,\text{TeV} & \text{FRB 200428} \\
		8\gamma_r^{-2}\,\text{PeV} & z=1.
	\end{cases}
\end{equation}

If the charge density be spread over a width $\ell \sim R/\gamma_{min} \sim
10^4\,$cm transverse to its direction of motion and radiation (a very oblate
shape), the maximum permitted by the condition that the fields add
coherently,
\begin{equation}
	\label{Rovergamma}
	E_e \ge
	\begin{cases}
		2\gamma_r^{-2}\,\text{GeV} & \text{FRB 200428} \\
		3\gamma_r^{-2}\,\text{TeV} & z=1.
	\end{cases}
\end{equation}
{These values are lower limits.  For the brighter FRB such as FRB 190523
\citep{R19} they are much greater than for the nominal 1 Jy FRB.  This
implies $\gamma_r \gg 1$ in order that $E_e$ not be impossibly large.}

The fact that FRB spectral structure typically consists of bands of width
$\Delta \omega \sim 0.1 \omega$ indicates that the radiating waves have a
minimum of $\sim 10$ periodically spaced charge peaks.  Individual regions
of unbalanced charge may have charges an order of magnitude less than
indicated by Eq.~\ref{Q}, with a corresponding reduction in $E_e$.  These
regions radiate coherently so the effective $Q$ is reduced in Eq.~\ref{Ee}
but not in Eqs.~\ref{dPdO} and \ref{dPdO2}.  This and the uncertain factor
$\gamma_r$ may make it possible to reconcile the values of
Eq.~\ref{Rovergamma} with the maximum electron energy $\sim 0.2$ TeV, above
which incoherent curvature radiation is energetic enough to make pairs in
the large magnetic field.  However, the much greater values of
Eq.~\ref{lambdabar} are probably excluded at cosmological distances.

If we set $\gamma_w = \gamma_{part}$ (so the radiating charges are
not a wave or pattern speed but are the actual particle speed) and use
Eq.~\ref{dPdO2} to determine $Q$ and Eq.~\ref{Ee} we find
\begin{equation}
	\gamma_{part} = \left({e \over \ell m_e c^2}\right)^{1/3}
	\left({\pi R^2 \over 4c}\left.{dP \over d\Omega}\right|_{obs}
	\right)^{1/6} \gamma_{min}^{-2/3}.
\end{equation}
Numerically
\begin{equation}
	\gamma_{part} \sim
	\begin{cases}
		300 ({10^4\,\text{cm} / \ell})^{1/3} &
		\text{FRB 20048} \\
		3500 ({10^4\,\text{cm} / \ell})^{1/3} & z=1.
	\end{cases}
\end{equation}
The corresponding $\gamma_r$ are $\sim 3 \times
(10^4\,\text{cm}/\ell)^{1/3}$ and $\sim 35 \times
(10^4\,\text{cm}/\ell)^{1/3}$, respectively.  The charges $Q$ may be found
from Eq.~\ref{Q}.
\subsection{Accelerating the Electrons}
\label{accelerate}
Can electrons be accelerated to the energies indicated Eqs.~\ref{lambdabar}
and \ref{Rovergamma}?  We calculate the required electric fields $E$ by
equating the power radiated by an electron in curvature radiation to the
power delivered by the electric field $\approx eEc$.  There are at least two
possible criteria:
\begin{enumerate}
	\item The power of the incoherent curvature radiation emitted by
		electrons with the energies given by Eq.~\ref{lambdabar} or
		\ref{Rovergamma}, the energies required for electrons to
		form bunches with the charges inferred from the observed
		radiation without being disrupted by electrostatic
		repulsion, must not exceed the power imparted by the
		accelerating electric field.  Their Lorentz factors
		$\gamma_{part}$ are generally much greater than
		$\gamma_{min}$.  The power an electron radiates as
		incoherent curvature radiation \citep{RL79}
\begin{equation}
	\label{Pcurv}
	P_{curve} = {2 \over 3}{e^2 \over c^3} a_\perp^2 \gamma_{part}^4
	\approx {2 \over 3} {e^2 \over c^3} {c^4 \over R^2} \gamma_{part}^4.
\end{equation}

For a ``bunch'' of charge $Q$ the elementary charge $e$ is replaced by $Q$
and $\gamma_{part}$ is replaced by $\gamma_w$ if the ``bunch'' is a
wave or pattern on an underlying particle distribution with different
		Lorentz factors.  Equating $P_{curve} = eEc$ \citep{KLB17},
\begin{equation}
	\label{Eincoheq}
	E \gtrsim {2 \over 3} {e \over R^2} \gamma_{part}^4,
\end{equation}
where $\gamma_{part} = Qe/\ell m_ec^2$ (Eq.~\ref{Ee}), is required.  The
resulting numerical values are shown in Table~\ref{Eincoh}.
\begin{table}
	\centering
	\begin{tabular}{|c|cc|}
		\hline
		$E$ (esu/cm$^2$) & FRB 200428 & $z=1$ \\
		\hline
		$\ell = \lambdabar$ & $3 \times 10^6 \gamma_r^{-8}$ &
		$2 \times 10^{19} \gamma_r^{-8}$ \\
		$\ell = R/\gamma_{min}$ & $3 \times 10^{-8}\gamma_r^{-8}$
		& $2 \times 10^5\gamma_r^{-8}$ \\
		\hline
	\end{tabular}
	\vskip 0.2in
	\begin{tabular}{|c|cc|}
		\hline
		$t$ (s) & FRB 200428 & $z=1$ \\
		\hline
		$\ell = \lambdabar$ & $2 \times 10^{-7}\gamma_r^7$ & $5 \times
		10^{-17} \gamma_r^7$ \\
		$\ell = R/\gamma_{min}$ & $6 \times 10^3\gamma_r^7$ & $1
		\times 10^{-6}\gamma_r^7$ \\
		\hline
	\end{tabular}
	\caption{\label{Eincoh} Minimum values of electric field (upper)
	(multiply by 300 to convert to V/cm) required to balance incoherent
	curvature radiation losses of electrons at the energies required to
	overcome Coulomb repulsion by radiating bunches and energy loss
	times (lower) if there is no accelerating field.  There is an
	additional criterion, that the electrons can be accelerated to the
	required energy (Eq.~\ref{Ee}) in a length $\lesssim R$, that sets a
	more stringent minimum of $E \gtrsim 5$ esu/cm$^2$ (1500 V/cm) for
	FRB 200428 if $\ell = R/\gamma_w$.}
\end{table}
Faraday's Law limits the electric fields that can be created by induction to
$E \lesssim B$, and vacuum breakdown \citep{HE36,S51,SY15,LK19} limits it to
$E \lesssim 2 \times 10^{12}\,$esu/cm$^2$.  The curvature radiation model
can be excluded as an explanation of ``cosmological'' FRB if $\ell \sim
\lambdabar$ unless $\gamma_r \gtrapprox 10$, but smaller values of
$\gamma_r$ are consistent with larger but possible values of $\ell$.

\item The electric field must replenish the
coherently radiated energy after the charge bunch has formed.  As shown in
Sec.~\ref{energetics}, the kinetic energies of the charge bunches are very
small, and must be replenished throughout a burst.  This criterion is
obtained from Eq.~\ref{Pcurv}, replacing $e$ by $Q$, using $\gamma_w =
100$ and the power delivered by the electric field $\approx QEc$:
\begin{equation}
	\label{Ecoheq}
	E \gtrsim {2 \over 3} {Q \over R^2} \gamma_w^4.
\end{equation}
The numerical results are shown in Table~\ref{Ecoh}, and are independent of
$\ell$ because the relevant Lorentz factor $\gamma_w$ is determined by
the observed frequency, not $\ell$.

\begin{table}
	\centering
	\begin{tabular}{c|cc|}
		\hline
		$E$ (esu/cm$^2$) & FRB 200428 & $z=1$ \\
		\hline
		All $\ell$ & $3 \times 10^6$ & $5 \times 10^9$ \\
		\hline
	\end{tabular}
	\caption{\label{Ecoh} Minimum values of electric field (multiply by
	300 to convert to V/cm) required to overcome coherent curvature
	radiation losses during the radiation of a charge bunch.  Because
	the relevant Lorentz factor is that of the coherent wave the results
	do not depend on the values of $\ell$ or of $\gamma_r$ that
	determine the minimum particle Lorentz factor.}
\end{table}
%


It may not be necessary that work done by the electric field continuously
replenish the kinetic energy of the coherently radiating charge bunches
(Table \ref{Ecoh}).  Energetic particles may be a sufficient energy
reservoir, intermittently producing charge bunches by plasma instability,
but if electrons cannot be accelerated to sufficient energy to form the
necessary charge bunches (as is the case for spherical bunches with $\ell
\sim \lambdabar$ and smaller $\gamma_r$) then sufficient coherent curvature
radiation cannot be emitted.
\end{enumerate}
\subsection{Origin of Accelerating Electric Field}
\label{origin}
Currents in a neutron star magnetosphere flow along closed magnetic loops,
anchored in the neutron star in analogy to Solar prominences, as in the
``magentar'' model of SGR.  A plasma instability may introduce a region of
large ``anomalous'' resistivity, much greater than the microscopic plasma
resistivity, interrupting the current flow and replacing the conductive
region with an effective capacitor.  Charge builds up on the boundaries of
the newly insulating region.

This is described as an $LC$ circuit with inductance $L \sim 4\pi r/c^2$ (in
Gaussian units), where $r$ is the radius of the current loop (that may be as
large as the magnetospheric radius $R$) and capacitance $C \sim A/
(4 \pi a)$, where $A$ is the cross-section of the current loop (that may be
as large as $\sim R^2$ for a distributed current) and $a$ is the width of
the gap that becomes insulating.  The charge on the surfaces of the gap
\begin{equation}
	Q_{gap}(t) = Q_0 \sin{t \over \sqrt{LC}} = \sqrt{LC} J_0 \sin{t
	\over \sqrt{LC}},
\end{equation}
where $t$ is the time since the insulating gap opened, $J_0$ was the
interrupted current, and $Q_0 = \sqrt{LC} J_0$.  For a distributed current
and a wide gap $A \sim r^2$ and $\sqrt{LC} \sim r/c \sqrt{r/a}$.  Then $J_0
\sim \Delta Brc/4\pi$, $Q_0 \sim \Delta B r^2\sqrt{r/a}/4\pi$, the voltage
drop $V \sim Q_0/C \sim \Delta B\sqrt{ra}$ and the electric field $E \sim
V/a \sim \Delta B\sqrt{r/a}$.  {When the current loop is interrupted the
required change in $B$ is $\Delta B \sim V/\sqrt{ra} \sim E\sqrt{a/r}$.} 
The fields indicated in the Tables for $\ell \sim R/\gamma_w$ can be
provided by plausible values of $\Delta B$.  {The potential drop $V \sim
\Delta B \sqrt{ra}$ must be $\ge E_e/e$.  Because the values of $r$, $a$ and
$\gamma_r$ (entering in Table~\ref{Eincoh}) are very uncertain it is not
possible to make numerical estimates.  The charges may be accelerated until
their incoherent or coherent radiative losses equal the power imparted by
the electric field, so that their energy density does not accumulate.}

The charges $Q_{gap}(t)$ are much larger than the radiating charges inferred
from Eq.~\ref{Q}, but are not moving relativistically and do not radiate
significantly.  Radiation will be emitted by the changing magnetic field.
On dimensional grounds, the expression for the power radiated in the dipole
approximation is roughly valid, where the dipole moment $\Delta \mu \sim
\Delta B r^3$, varies on a characteristic time scale $\sim 1/\omega \sim
c/r$ and $r$ is the radius or characteristic size of the loop:
\begin{equation}
	\label{dipolerad}
	P \sim {(\Delta \mu)^2 \omega^4 \over 3 c^3} \sim {(\Delta B)^2 c^3
	\over 3 \omega^2}.
\end{equation}
For the maximum plausible $\Delta B \sim 10^{15}$ gauss and the observed
FRB L-band frequency, $P \sim 10^{41}\,$ergs/s and would be unbeamed, in
contradiction to the argument of Sec.~\ref{Lorentz} for FRB 200428.  Such
unbeamed power would be insufficient to power ``cosmological'' FRB.  Hence
the radiation of Eq.~\ref{dipolerad} is unlikely to be related to observed
FRB.

The achievable value of $E$ may be limited by breakdown creation of
electron-positron pairs, either the Schwinger vacuum breakdown that occurs
for $E \gtrapprox 2 \times 10^{12}\,$esu/cm$^2$, or the curvature
radiation-driven pair production cascade breakdown believed to occur in
pulsars.  Even if breakdown occurs, it may not necessarily ``short out'' the
electric field and accumulated charges because the region of breakdown may
still be resistive as a result of plasma instability.  If the current loop
is wide ($\ell \sim R/\gamma_w$), $E$ may be large enough to accelerate the
electrons to the energies necessary to overcome Coulomb repulsion.  Each
portion of the area $A$ accumulates charge, limited independently by 
breakdown in the capacitive gap, so that it may be possible to produce the
necessary thin sheet charge distribution.

Faraday's law
\begin{equation}
	\label{Faraday}
	\nabla \times {\vec E} = -{1 \over c}{\partial {\vec B} \over
	\partial t}
\end{equation} implies
\begin{equation}
	{E \over \Delta x} \sim {1 \over c} {\Delta B \over \Delta t}.
\end{equation}
$\Delta B \le B$ (defining $B$ as its maximum magnitude).  Causality
requires $\Delta t \ge \Delta x/c$ so that
\begin{equation}
E \lesssim \Delta B \le B.
\end{equation}
This is a general limit on the electric fields that can be produced in
a relaxing current-carrying magnetosphere.

Changing the magnetic field within a loop of area $r^2$ by $\Delta B$ in a
time $\tau$ produces an inductive electromotive force (EMF) 
\begin{equation}
	\label{Vinduct}
	\begin{split}
		V_{inductive} &\sim {r^2 \Delta B \over c\tau} \\ &\sim 3
		\times 10^{10} {\Delta B \over 10^8\,\text{gauss}}{r^2 \over
		10^{12}\,\text{cm}^2} {0.1\,\text{s} \over \tau}\,
		\text{esu/cm}
	\end{split}
\end{equation}
and an electron energy
\begin{equation}
	\label{Eeinduct}
	\begin{split}
		E_e &= e V_{inductive} \\
		& \sim 10 {\Delta B \over 10^8\,\text{gauss}}{r^2 \over
		10^{12}\,\text{cm}^2} {0.1\,\text{s} \over \tau}\,\text{TeV}.
	\end{split}
\end{equation}

In FRB 200428 the EMF required to accelerate particles to the minimum energy
for $\ell = \lambdabar$ and $\gamma_r = 1$ (Eq.~\ref{lambdabar}) can be
provided by $\Delta B \sim 10^8$ gauss if the loop encompasses much of the
magnetosphere ($R \sim 10^6\,$cm) and if $\tau \sim 0.1\,$s, as observed for
SGR.  If $\ell = R/\gamma_w$ and $\gamma_r = 1$ (Eq.~\ref{Rovergamma}),
$\Delta B \sim 2 \times 10^4$ gauss would be sufficient.  For a nominal 1
Jy-ms FRB at $z=1$, $\ell \sim \lambdabar$ and $\gamma_r = 1$ would require
$\Delta B \sim 10^{11}\,$gauss but $\ell \sim R/\gamma_w$ and $\gamma_r = 1$
would only require $\Delta B \sim 3 \times 10^7\,$gauss.  Without a detailed
understanding of the magnetohydrodynamics and plasma physics of SGR activity
we cannot decide if these values are plausible, but they violate no physical
law.
\subsection{Energetics}
\label{energetics}
The magnetic energy dissipated is obtained using Eq.~\ref{Vinduct} and $r
\sim R$ to obtain the minimum $\Delta B$ required to accelerate electrons to
the energy $E_e = V_{inductive}e$:
\begin{equation}
	\label{energy}
	\begin{split}
		{\cal E} & \sim {1 \over 3} B \Delta B R^3 \sim {B c \tau R Q
		\over 3 \ell} \\ & \sim
	\begin{cases} 
		3 \times 10^{39}\,\text{ergs} & \text{FRB 200428} \\
		3 \times 10^{43}\,\text{ergs} & z=1,
	\end{cases}
	\end{split}
\end{equation}
where the numerical values assume $\ell \sim \lambdabar$ (larger $\ell$
would lead to lesser values), $\gamma_r = 1$ and the observed width of FRB
outbursts $\tau \sim 0.1\,$s; for FRB 200428 $B = 2 \times 10^{14}\,$gauss
\citep{I16} and for a burst at $z=1$ $B = 10^{15}\,$gauss have been
assumed.  The value of $\cal E$ for FRB 200428 is consistent with the
observed X-ray fluences of SGR 1935$+$2154.  For ``cosmological'' FRB the
value of $\cal E$ is consistent with giant outbursts of Galactic SGR, but
the argument of Sec.~\ref{180620} indicates that only less powerful SGR
outbursts may produce FRB.

Eq.~\ref{lambdabar} ($\ell = \lambdabar$) would permit $\sim 10^6$ bursts in
the lifetime of SGR 1934$+$2154 and $\sim 10^4$ repetitions for the nominal
``cosmological'' FRB if $B \sim 10^{15}$ gauss.  The number of repetitions
could be several thousand times greater if $\ell = R/\gamma_w$
(Eq.~\ref{Rovergamma}).  These values are obtained from the required
inductive EMF, not directly from the change in magnetostatic energy.  If the
magnetic field is regenerated from internal motions, there could be yet more
repetitions.  Weaker ``cosmological'' bursts, such as those of FRB 121102,
require smaller $Q$, $E_e$, $V_{inductive}$, and $\Delta B$, and could
repeat many more times during the active lifetimes of their sources.

The electric fields within the charge bunches
\begin{equation}
	\label{E}
	\begin{split}
		E & \sim {Q \over \ell^2} \\ & \sim
	\begin{cases}
		5 \times 10^2 (R/100\ell)^2\gamma_r^{-2}\,\text{esu/cm}^2
		& \text{FRB 200428} \\
		8 \times 10^5 (R/100\ell)^2\gamma_r^{-2}\,\text{esu/cm}^2
		& z=1.
	\end{cases}
	\end{split}
\end{equation}
If $\ell \sim \lambdabar$ and $\gamma_r \sim 1$ the field estimated for the
cosmological FRB exceeds the Schwinger pair-production vacuum breakdown
field \citep{HE36,S51,SY15} several-fold.  This paradox is resolved if the
charge distribution is oblate, with $\ell \gg \lambdabar$, or if $\gamma_r
\gg 1$.  It might seem unlikely that charge would be concentrated into thin
sheets perpendicular to its direction of motion and the magnetic field
lines, but there is a strong selection effect favoring the observation of
such emitting geometry because for it the fields add coherently, making the
radiation stronger and more observable.

The kinetic energies of the motion of the net charges $Q$ (Eqs.~\ref{Q},
\ref{Ee}) are very small, $\sim 8 \times 10^{20}$ ergs for FRB 200428 and
$\sim 2 \times 10^{27}$ ergs for the nominal FRB at $z=1$ even if $\gamma_r
= 1$.  {The energy driving the FRB, derived from magnetostatic energy,
may flow directly to the radiating charges via the electric fields of
Sec.~\ref{origin}, continually replenishing the kinetic energies of the
radiating charges.  The radiating region need not be quasi-neutral.}
\subsection{Curvature Radiation {\it vs.\/} Compton Scattering}
The relativistic electrons emitting curvature radiation are moving in the
soft gamma-ray radiation field of the SGR.  It is necessary to compare the
power they emit in curvature radiation to their energy loss by Compton
scattering.  If the latter were to dominate, then it would be difficult to
accelerate a population of electrons to the energies necessary to emit a
FRB.  {It is only possible to estimate the lower bound resulting from
Compton scattering by the electrons contributing to the emission of coherent
(FRB) curvature radiation.  The density of any quasi-neutral plasma is
unknown, and its Compton scattering power cannot be estimated.  Hence the
following is only a demonstration that Compton scattering need not be a
catastrophic energy loss.}

The power the electrons lose to Compton scattering is
\begin{equation}
	\label{Compt}
	P_{Compt} \approx n_\gamma N_e \sigma_{KN} E_e c,
\end{equation}
where
\begin{equation}
	\label{ngamma}
	n_\gamma \sim {L_{SGR} \over 4 \pi R^2 h\nu_\gamma c}
\end{equation}
is the number density of soft gamma-rays, $N_e = Q/e$ is the number of
electrons in the charge bunch, $\sigma_{KN} \approx \pi r_e^2 \ln{(2h
\nu_\gamma E_e/m_e^2c^4)}/(h\nu_\gamma E_e/m_e^2c^4)$ is the Klein-Nishina
cross-section ($r_e = e^2/m_ec^2$ is the classical electron radius) and
$E_e$ is the electron energy.  In this regime of highly relativistic
electrons scattering soft gamma-rays, nearly the entire electron kinetic
energy is lost to the photon in a single scattering.

For FRB 200428, using Eqs.~\ref{Q}, \ref{Ee} and \ref{ngamma}, $L_{SGR} \sim
6 \times 10^{39}$ ergs/s (at 6 kpc distance), $h\nu_\gamma \sim 50$ keV
\citep{M20} and $\gamma_w = 100$, Eqs.~\ref{Pcurv} and \ref{Compt} yield
\begin{equation}
	\label{ratio}
	\begin{split}
		{P_{curve} \over P_{Compt}} &\sim {8 \over 3} {Q c
		(h\nu_\gamma)^2 \gamma_w^4 \over L_{SGR} e^3
		\ln{(2h\nu_\gamma E_e/m_e^2c^4})} \\
		& \sim 300 {6 \times 10^{39}\,\text{ergs/s} \over L_{SGR}}.
	\end{split}
\end{equation}
This value is uncertain, but is consistent with the assumption that
Compton scattering losses do not exceed the radiated power and therefore the
validity of Eq.~\ref{Ecoheq} as a condition on the electric field.  The use in
Eq.~\ref{Compt} of the lower bound Eq.~\ref{Ee} on $E_e$ is balanced, except
for the slowly varying logarithm, by the energy dependence of $\sigma_{KN}$.

Despite the intense soft gamma-ray radiation field, the quadratic dependence
of the coherent $P_{curve}$ on $Q$ makes it possible for it to exceed
$P_{Compt}$ that is only proportional to one power of $Q = N_e e$.  An
additional factor of $Q$ enters $P_{Compt}$ through the minimum electron
energy (Eq.~\ref{Ee}), but this is nearly cancelled by the inverse energy
dependence of the Klein-Nishina cross-section.  The number of coherently
radiating charges in the bunch or wave $N_e = Q/e$ is $\sim 10^{20}$ for FRB
200428 and $\sim 10^{23}$ for the cosmological FRB.  These enormous values
and the quadratic dependence on $Q$ (or $N_e$) that makes the FRB bright
enough to observe also make Compton losses comparatively unimportant.
\subsection{Why Not SGR 1806$-$20}
\label{180620}
The strongest argument against the SGR-AXP hypothesis was empirical:
During an unrelated observation, the giant 27 December 2004 outburst of SGR
1806$-$20 was in a radio telescope sidelobe but no FRB was detected
\citep{TKP16}.  Although the sidelobe had sensitivity about 70 dB less
than that of the main beam, the fact that the SGR was $\sim 3 \times 10^5$
times closer than a typical ``cosmological'' FRB, as well as the
extraordinary brightness of the SGR, led to an upper limit on the ratio of
the radio to soft gamma-ray fluences of $< 10^7$ Jy-ms/(erg/cm$^2$).  This
is more than five orders of magnitude less than the observed fluence ratio
$> 2 \times 10^{12}$ Jy-ms/(erg/cm$^2$) of FRB 200428/SGR 1935$+$2154.

There are at least three possible explanations.
\begin{enumerate}
	\item Compton energy loss (Eq.~\ref{ratio}).  The soft gamma-ray
		luminosity of SGR 1806$-$20 during its giant outburst
		\citep{P05} was more than seven orders of magnitude greater
		than that of SGR 1935$+$2154 during FRB 200428; this was
		only partially offset by a value of $h\nu_\gamma$ less than
		two orders of magnitude greater, leading to a ratio
		$P_{curve}/P_{Compt} \sim 10^{-3}$ for a burst like FRB
		200428.  Emission of GHz curvature radiation by SGR 1806-20
		was suppressed by Compton scattering energy losses of the
		required relativistic electrons.

At intensities greater than {$\sim 10^{29}$} ergs/cm$^2$-s (luminosities
$\gtrsim 10^{42}$ ergs/s for an isotropically emitting neutron star)
radiation and energetic particles thermalize to black-body equilibrium by
processes that turn two incoming particles into three outgoing particles:
radiative Compton scattering, three photon pair annihilation \citep{K96} and
photon splitting in a strong magnetic field.  The result is an opaque
equilibrium photon-pair plasma in which relativistic particles suffer
runaway Compton and Coulomb scattering energy loss and radio radiation
cannot propagate.
\item Observations of FRB 200428 \citep{L20} indicate that the observable
FRB/SGR flux ratio may vary from burst to burst by at least eight orders of
magnitude, plausibly because of beaming (Sec.~\ref{Lorentz}).
\item FRB radiation, unlike SGR, may be narrowly collimated, as suggested by
	Eq.~\ref{dPdO}.  This is discussed in Sec.~\ref{secFvN}.
\end{enumerate}
\subsection{Distribution of FRB Fluxes}
\label{secFvN}
{If the beams of particle ``bunches'' emitting FRB are randomly
distributed in direction, then the distribution of FRB fluxes can be
obtained from Eq.~\ref{dPdO}.  Results are shown in Fig.~\ref{FvN}.  The
logarithmic slope is about $-1/4$; for the corresponding relation between
frequency and fluence the logarithmic slope is about $-1/3$ because the
brighter bursts with larger Doppler factors are shorter.

\begin{figure}
	\centering
	\includegraphics[width=0.95\columnwidth]{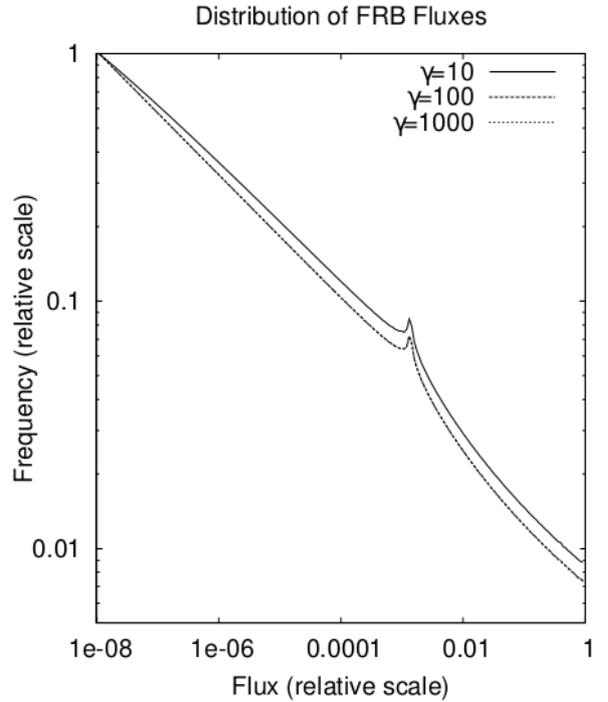}
	\caption{\label{FvN} Frequency distribution of FRB fluxes emitted by
	randomly oriented, narrowly collimated, radiating charge
	``bunches''.  The peak around ${\cal F} \approx 10^{-3}$ arises
	because Eq.~\ref{dPdO} is stationary around $(\gamma\theta)^2 = 2$
	and $\cos{2\phi} = 1$.  The distribution is almost independent of
	$\gamma$ for $\gamma \gg 1$ (the curves for $\gamma = 100$ and
	$\gamma = 1000$ are indistinguishable).}
\end{figure}

The distribution of observed fluxes is given by the convolution of the
distribution shown in the Figure with the unknown distribution of intrinsic
burst strengths.  Nonetheless, this is consistent with the fact that FRB
200428 was discovered in $\sim 10^7\,$s of on-source observing time
\citep{B20,C20}, while a burst of $\sim 4 \times 10^{-8}$ of its intensity
was discovered in a 62 minute observing run \citep{FAST}, $\sim 3 \times
10^{-4}$ that long.  It is also consistent with the comparative frequency of
strong (10--100 times the median) bursts from FRB 180916.J0158$+$65
\citep{Ma20}.

``Superbursts'' like FRB 200428 are produced when a narrowly collimated beam
is directed towards the observer \citep{K17a}.  They are unusual, but not
rare, as indicated by the low slope in Fig.~\ref{FvN}, and represent most of
the time-integrated FRB power.  The mean FRB power emitted is several orders
of magnitude less than the isotropic-equivalent power of a superburst, or of
any observed burst, because of collimation.  ``Cosmological'' FRB may be
superbursts of emitters whose mean FRB power is much less than their
isotropic-equivalent power.  This is also implied by their low duty factor
\citep{K18a}.}
\section{Discussion}
The discovery and identification of FRB 200428 resolved the first question
about FRB: What astronomical objects produce them?  It took 13 years from
their discovery (and 7 years from the time their reality became generally
accepted) to answer this question because of the difficulty of accurate
localization.  The similar difficulty of localizing gamma-ray bursts meant
that their identification took 25 years, as did the recognition of
extra-Galactic radio sources as the products of active galactic nuclei
(AGN).

Identification of FRB with rotating neutron stars predicts that FRB activity
should be modulated, at some level, at the rotation period.  Periodicity has
not been observed in FRB 121102, the only FRB for which abundant data exist
\citep{Z18}; see discussion in \citet{K19}.  If ``cosmological'' and
Galactic FRB are qualitatively similar phenomena, periodicity should be
detectable in any FRB that repeats frequently.  Periodicity will be easier
to detect in FRB identified with Galactic SGR because their periods would be
known {\it a priori\/} from gamma-ray observations of the SGR/AXP.

The magnetospheric densities implied by Eq.~\ref{Q} and the constraint on
the dimensions of a radiating charge bunch $< R/\gamma_w$ exceed the
critical plasma density at observed FRB frequencies for the parameters of
cosmological FRB.  However, this limit on propagation is inapplicable.  The
plasma is strongly magnetized (so strongly that the electrons' motion
transverse to the field, the direction of the electric vector of a
transverse wave propagating along the field, is quantized).  In addition, the
electrons' longitudinal motion is highly relativistic (Eq.~\ref{Ee}),
increasing their effective mass by the factor $\gamma_{part}$.  Finally, the
radiating charge bunches may be confined to a shell thinner than the skin
depth, like the currents in a metallic antenna radiating radio waves.
Propagation and escape of the radiation are beyond the scope of this paper,
but are issues that must be faced by any model in which FRB are emitted from
a compact region, as required by their narrow temporal structure.

Identification of FRB with SGR does not itself explain their mechanism.
Their high brightness temperatures require coherent emission, but there is
no understanding of their charge bunching.  Even in pulsars, discovered 53
years ago, the mechanism of charge bunching remains uncertain.  Acceleration
of relativistic particles is nearly ubiquitous in astrophysics \citep{K91},
and is also required to explain FRB, but is not understood from first
principles; if we had not inferred it from observations in AGN, Solar
activity, supernova remnants, pulsars, FRB and many other phenomena, we
would not have predicted it.

The presence of an intense thermal (X-ray and soft gamma-ray) radiation
field interferes with the acceleration and propagation of relativistic
electrons.  At sufficiently high radiation energy densities, radiative and
particle energy thermalizes to a dense equilibrium pair-photon plasma
\citep{K96}.  This predicts that SGR with luminosities $\gtrsim 10^{42}$
ergs/s do not make FRB comparable to FRB 200428.

The issues discussed here of the radiating charges $Q$ and their implied
electric fields extend beyond curvature radiation models, and apply however
the charges are bunched, whether by plasma instability, maser amplification,
or another mechanism.  In any model, radiation can only be produced by
accelerated charges or changing currents.  It is difficult to produce
beaming from changing currents because conservation of charge and the
assumption of quasi-neutrality imply that current is constant along bundles
of field lines; a relativistically moving current front cannot be produced
without creating net charge density.  The required $Q$ are determined by the
very general Eq.~\ref{Q} and the particle Lorentz factors by Eq.~\ref{Ee}
that are not specific to curvature radiation.  This does not exclude sources
outside an inner neutron star magnetosphere, but Eq.~\ref{dPdO} applies and
smaller $Q$ imply larger $\gamma_w$, narrower beaming and, if
$\gamma_w$ is the Lorentz factor of an actual particle bunch, higher
particle energy.


{The extreme sensitivity of $\cal F$ and ${\cal F}_{obs}$ to $\theta$
(Eqs.~\ref{dPdO} and \ref{dPdO2}) may explain the excess \citep{K17b} of very
bright bursts; not enough bursts have been observed to sample fully the
distribution of source strengths that otherwise would follow the
$d\ln{N}/d\ln{{\cal F}_{obs}} = -3/2$ distribution of sources homogeneously
distributed in Euclidean space.  This also suggests that repeating FRB may
show ``superbursts'' exceeding their other observed bursts by orders of
magnitude (Sec.~\ref{secFvN}).  FRB 200428 was apparently such a superburst
of SGR 1935$+$2154.}

{\cite{Y20} reported that the
outburst of SGR 1935$+$2154 coincident with FRB 200428 had an X-ray spectrum
distinct from that of its other outbursts, with a much higher energy cutoff.
This is consistent with relativistic bremsstrahlung emitted by the
FRB-emitting electrons, narrowly collimated towards the observer (parallel
to the coherent radio frequency radiation), in addition to the roughly
isotropic SGR radiation.  Such relativistic bremsstrahlung would extend to
photon energies $E_e$ (Eq.~\ref{Ee}), but its intensity cannot be predicted
from the radio intensity because they scale as different powers of $Q$.}

Of the 30 SGR and AXP (the quiescent counterparts of SGR) and candidates now
known \citep{OK14}, only SGR 1935$+$2154 has been detected as a source of
FRB.  What distinguishes SGR 1935$+$2154?  Its spin period of 3.245 is near
the short end of the distribution of spin periods, but it is not the
shortest.  Its inferred (from spindown, noting that SGR/AXP spindown rates 
vary by factors ${\cal O}(1)$) surface magnetic field of $2.2 \times
10^{14}$ gauss is well within the range of the fields of other SGR/AXP.  I
suggest that it is distinguished only by having a storm of activity during
the short interval during which the wide field instruments CHIME/FRB and
(especially) STARE2 have been operating.  This explanation leads to the
prediction that other SGR produce observable FRB during their storms of
activity.

This prediction could be tested, and exploited if found to be correct,
by staring at the positions of known SGR/AXP with small dedicated
telescopes.  The $d = 4.5$ m diameter L-band telescopes in the DSA system
\citep{DSA} would be suitable.  Their ideal antenna gains ($\approx \pi^2
d^2/\lambda^2$) are about 38 dB, a great improvement in sensitivity over
STARE2 that could not detect the FRB of SGR 1935$+$2154 other than its
first, MJy-ms, giant burst.  Staring gives a greater likelihood that a
source will be within the field of view than CHIME/FRB \citep{CF}, whose
instantaneous field of view is about 0.5\% of the sky.
\section*{Acknowledgements}
I thank Wenbin Lu for many useful discussions and NSF AST 84-12895 for the
HP 11C calculator used in this work.
\section*{Data Availability}
This is a theoretical paper that does not involve any new data.

\bsp 
\label{lastpage} 
\end{document}